\documentstyle[amssymb,aps,prl,multicol,epsfig]{revtex}

\begin{document}

\title{Why Ni$_3$Al is an itinerant ferromagnet but Ni$_3$Ga is not}
\author{A. Aguayo$^1$, I.I. Mazin and D.J. Singh}
\address{Center for Computational Materials Science, Naval Research
Laboratory,
Washington, DC 20375 \\
$^1$also at School of Computational Sciences, George
Mason University, Fairfax, VA 22030}
\date{\today }

\maketitle
\begin{abstract}
Ni$_3$Al and Ni$_3$Ga are closely related materials on opposite sides
of a ferromagnetic quantum critical point.
The Stoner factor of Ni is virtually the same in both compounds and
the density of states is larger in Ni$_3$Ga. So, according
to the Stoner theory, it should be more magnetic, and, in LDA calculations,
it is. However, experimentally,
it is a paramagnet, while Ni$_3$Al is an itinerant ferromagnet.
We show that the critical spin fluctuations are
stronger than in Ni$_3$Ga, due to a weaker q-dependence of the
susceptibility, and this effect is strong enough to reverse the trend. The
approach combines LDA calculations with
the Landau theory and the fluctuation-dissipation theorem using
the same  momentum cut-off for both compounds.
The calculations provide evidence for strong, beyond LDA, spin
fluctuations associated with the critical point in both materials,
but stronger in Ni$_3$Ga than in Ni$_3$Al.
\end{abstract}


\begin{multicols}{2}

Recent low temperature experiments on clean materials near
ferromagnetic quantum critical points (QCP) have revealed a remarkable
range of unusual properties, including non-Fermi liquid scalings over 
a large phase space, unusual transport, and
novel quantum ground states, particularly coexisting ferromagnetism
and superconductivity in some materials.
Although criticality usually implies a certain 
universality,
present experiments show considerable material
dependent aspects that are not well
understood, \cite{laughlin}
{\em e.g.} the differences between
UGe$_2$ and URhGe \cite{saxena,aoki}
and ZrZn$_2$, \cite{pfleiderer}
which both show coexisting ferromagnetism and
superconductivity but very different phase diagrams,
in contrast to MnSi, where very clean samples show no hint of
superconductivity around the QCP.
\cite{pf2}
Generally, approaches based on density functional theory (DFT)
are successful in accounting for material dependence in cases
where sufficiently accurate approximations exist.
To proceed in this direction it is
useful to study benchmark systems for which detailed experimental
data are available and which pose challenges to theory. Here we
report a study of the closely related compounds
Ni$_3$Al and Ni$_3$Ga. Both of these have the ideal
cubic Cu$_3$Au $cP4$ structure, with very similar lattice constants,
$a=3.568$ \AA ~and $a=3.576$ \AA, respectively,
and have been extensively studied by various
experimental techniques. Ni$_3$Al is a weak
itinerant ferromagnet,
$T_{c}$ = 41.5 K and magnetization, $M$=0.23 $\mu _{B}$/cell
(0.077 $\mu _{B}$/Ni atom) \cite{boer}
with a QCP under pressure
at $P_c$=8.1 GPa,
\cite{niklowitz}
while Ni$_3$Ga is a strongly renormalized paramagnet\cite{hayden}.
Further, it was recently reported that
Ni$_3$Al shows non-Fermi liquid transport over
a large range of $P$ and $T$ range down to very low $T$.
\cite{steiner}

DFT is an exact ground state theory, and as such should properly describe
the magnetic ground states of metals. However, common approximations
to DFT, such as the LDA and GGA, are based on the properties of
the uniform electron gas at densities that occur in solids. At these densities
it is rather stiff with respect to spin fluctuations and
 is not close to any magnetic instability. As a result, the
LDA description of magnetism is at a quasi-classical mean field level,
({\em i.e.} Stoner level), and neglects fluctuations due to
 soft magnetic degrees of freedom. This leads to misplacement
of QCPs and overestimates of the magnetic tendencies
of materials near QCPs, as well as such known
problems as the incorrect description of singlet states in
molecules with magnetic ions. In fact,
practically all cases where the LDA substantially
overestimates the tendency towards magnetism are materials near a QCP
\cite{sr-327,zrzn2,sc3in,nac2} -- a fact that can potentially be used
as a screen for materials with large fluctuation effects.
\cite{hubbard}
Previous LDA calculations showed that the
magnetic tendency of both materials is overestimated within the
LDA, and that Ni$_3$Ga is incorrectly predicted to be a ferromagnet.
\cite{buiting,moruzzi1,min,xu,guo,hsu}
Moreover, as our present results show, in the LDA the tendency to magnetism 
is stronger in Ni$_3$Ga than Ni$_3$Al,
{\em opposite to the experimental trend}.
This poses an additional challenge to any theory striving to 
describe the material dependent aspects of quantum criticality.
The two materials are expected to be very similar electronically
(our results confirm this, and identify the small difference between
the two as due to relativistic effects associated with Ga in Ni$_3$Ga).
Thus they offer
a very useful and sensitive benchmark for theoretical approaches. We use this
to test an approach based on the fluctuation dissipation theorem
applied to the LDA band structures with an ansatz for the cut-off
$q_c$. We find that this approach corrects the ordering of the
magnetic tendencies of the materials, and gives the right ground
states at ambient pressure as well as a reasonable value of $P_c$
for Ni$_3$Al.

\begin{figure}[tbp]
\centerline{\epsfig{file=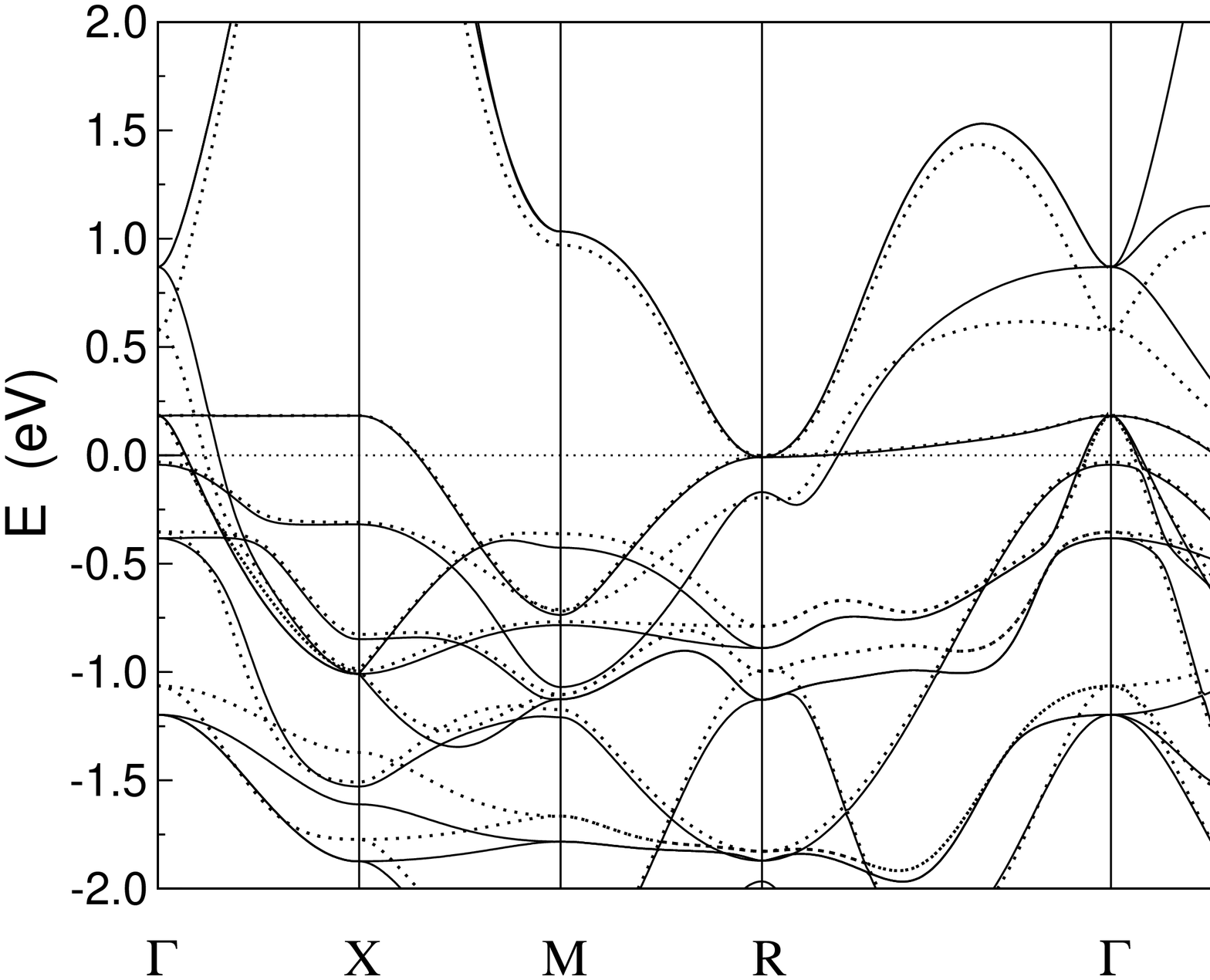,width=0.75\columnwidth}} 
\centerline{\epsfig{file=nialga2z.eps,width=0.90\columnwidth}  }
\caption{Calculated LDA band structure (top)
and density of states (bottom) per f.u.~for
non-spin-polarized Ni$_{3}$Al (solid lines) and Ni$_{3}$Ga (dotted lines).
$E_F$ is at 0 eV.  }
\label{bands} \end{figure}

\begin{figure}[tbp]
\epsfig{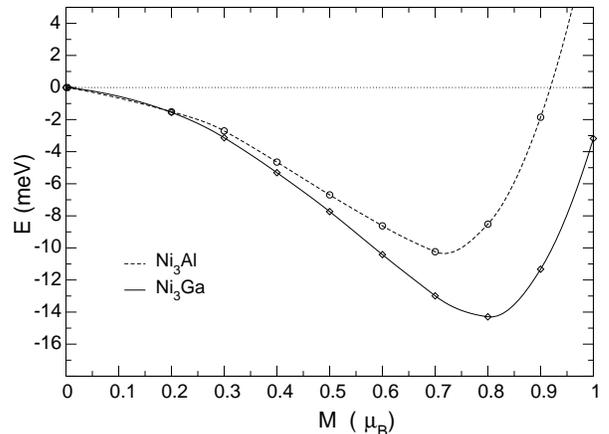}
\caption{Energy {\em vs.} fixed spin moment for Ni$_3$Al and Ni$_3$Ga
at the experimental lattice parameters.
The energy zero is set to the non-spin-polarized value.}
\label{FSM} \end{figure}

\begin{figure}
\hskip -.8cm \epsfig{file=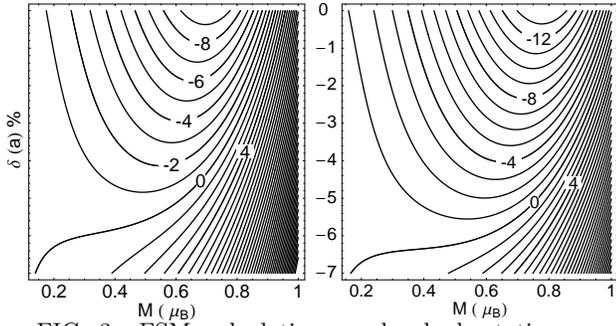,width=1.07\columnwidth}
\caption{FSM calculations under hydrostatic pressures. Magnetic energy,
defined as the energy relative to the non-spin-polarized result
at the same volume,
as a function of the moment and linear compression.
Left and right panels correspond to Ni$_3$Al and
Ni$_3$Ga, respectively.}
\label{P}
\end{figure}

Our LDA calculations were done using the
general potential linearized augmented-plane-wave (LAPW) method
with local orbital extensions
\cite{singh-book,singh-lo}
in two implementations \cite{singh-lo,wei,WIEN}, with the exchange-correlation
functional of Hedin-Lundqvist with the von Barth-Hedin
spin scaling \cite{hl,hl2}.
Up to 816 inequivalent \textbf{k-}points 
were used in the self-consistent calculations,
with an LAPW
basis set defined by the cut-off $R_{S}K_{max}$=9, plus
local orbitals to relax linearization errors.
Larger numbers of
\textbf{k-}points between 2300 and 4060 were
used in the Fermi surface integrations.
The LDA electronic structure is given in Fig. \ref{bands} and Table 1, while results of fixed
spin moment calculations of the magnetic properties at the experimental
lattice parameters and under hydrostatic compression are given in
Figs. \ref{FSM} and \ref{P}.
The two compounds are
very similar in both electronic and magnetic properties, the main
apparent difference being the higher equilibrium moment of Ni$_3$Ga
(0.79 $\mu_B$/f.u. {\em vs.} 0.71 $\mu_B$/f.u.),
in agreement with other full potential calculations. \cite{guo,hsu}

The propensity
towards magnetism may be described in terms of the Stoner criterion, 
$IN(E_{F})$, where $I$ is the so-called Stoner parameter, which derives
from
Hund's rule coupling on the atoms. For finite magnetizations,
the so-called extended Stoner model \cite{Krasko}, states
that to the second order in the spin density the magnetic stabilization
energy is expressed as
$\Delta E={M^{2}}[\int_0^Mm~dm/2\tilde{N}(m)-I/4]$, 
where $\tilde{N}(M)$ is the density of states averaged over the exchange
splitting corresponding to the magnetization $M.$
Fitting our fixed spin moment results to this expression, 
we find $I_{Al}=0.385$ eV and $I_{Ga}=0.363$ eV. These gives
$IN(E_F)=$1.21
and $IN(E_F)$ = 1.25 for Ni$_{3}$Al and Ni$_{3}$Ga, respectively.
Both numbers are larger than 1, corresponding to a ferromagnetic instability,
and the value for Ni$_{3}$Ga is larger than that for Ni$_{3}$Al.
Importantly, the difference comes from the density of states, since  $%
I_{Al}>I_{Ga}.$  
In both compounds, magnetism is suppressed by
compression, with an LDA critical point at a value $\delta a/a \sim$
-0.05 -- -0.06. In Ni$_3$Al, the critical point at $\delta a/a$ =-0.058
corresponds to the pressure of $P_c=$50 GPa, \cite{Pnote} which is
much higher than the experimental value.
It is interesting that, as in ZrZn$_2$ \cite{zrzn2},
the exchange splitting is very strongly {\bf k}-dependent;
for instance, in Ni$_3$Al at some points it is as small as 40 meV/$\mu_B$
near the Fermi level, while at the others (of pure Ni d character) it is
close to 220 meV/$\mu_B$.

Notwithstanding
the general similarity of the two compounds,
there is one important difference near the Fermi level,
specifically, the light band crossing the Fermi level in the middle
of the $\Gamma $-M or $\Gamma $-X directions is
steeper in Ni$_{3}$Al (Fig. \ref{bands}). 
This,
in turn, leads to smaller density of states. This comes from
a different position of the top of this band at the $%
\Gamma $ point, 0.56 eV in Ni$_{3}$Ga and 0.85 eV in Ni$_{3}$Al. The
corresponding electronic state is a mixture of Ni $p$ and Al (Ga) $p$
states, and is the only state near the Fermi level with substantial Al (Ga)
content. Due to relativistic effects, the Ga $p$ level is lower than the Al $%
p$ level and this leads to the difference in the position of the
corresponding hybridized state. Note that this is a purely scalar relativistic
effect. We checked that including spin orbit does not produce any further
discernible difference.

Returning to magnetism, the
fixed spin moment calculations provide the energy $E$ as a function of the
magnetization $M$
(Fig. \ref{FSM}).
One can write a Landau expansion for $E(M)$ as

\begin{equation}
E(M) = a_2M^2/2+a_4M^4/4+a_6M^6/6 + \cdots
 \label{EM}
\end{equation}

Treating this as a mean-field expression and adding the effects of spin
fluctuations \cite{shimizu} leads to renormalization of the expansion
coefficients.
The renormalized coefficients $\tilde{a}_{i}$ are
written as power series in the averaged square of the magnetic moment
fluctuations beyond the LDA, 
$\xi ^{2}$,

\begin{equation}
\tilde{a}_{2n}=\sum_{i\ge 0}C_{n+i-1}^{n-1}a_{2(n+i)}\xi
^{2}\prod_{k=n}^{n+i-1}(1+2k/3).  \label{corr}
\end{equation}

$\xi $ may be estimated by requiring that the
corrected Landau functional (\ref{corr}) reproduces the experimental magnetic
moment (for Ni$_{3}$Al) or experimental magnetic susceptibility (for Ni$_{3}$%
Ga). The \textquotedblleft experimental\textquotedblright\ $\xi $'s
obtained in this manner are
are 0.47 and 0.55, respectively, which implies that spin fluctuation
effects must be stronger in Ni$_{3}$Ga than in Ni$_{3}$Al.

We shall now make a link between this fact and the electronic structures.
A standard formula for estimating $\xi ^{2}$ comes from
the fluctuation-dissipation theorem \cite{moriya}, which establishes that

\begin{equation}
\xi ^{2}=(2\hbar /\Omega )\int d^{3}q\int (d\omega/2\pi )
\mathrm{Im}\chi (\mathbf{q},\omega ), \label{FDT}
\end{equation}
where $\Omega $ is the Brillouin zone volume, and $\chi$ is the magnetic
susceptibility.
Using the lowest order expansion for $\chi$,
\begin{eqnarray}
\chi _{0}(\mathbf{q},\omega ) &=&N(E_{F})-aq^{2}+ib\omega /q  \label{ab} \\
\chi ^{-1}(\mathbf{q},\omega ) &=&\chi _{0}^{-1}(\mathbf{q},\omega )- {\em I},
\end{eqnarray}%
where $\chi _{0}(\mathbf{q},\omega )$ is the non-interacting susceptibility,
one can derive a formula for $\xi^2$ \cite{shimizu,moriya},
whose coefficients can be related to the characteristics of the electronic structure 
\cite{moriya,larson}.
The final results reads

\begin{equation}
\xi ^{2}=\frac{bv_{F}^{2}N(E_{F})^{2}}{2a^{2}\Omega }[Q^{4}\ln
(1+Q^{-4})+\ln (1+Q^{4})],
\end{equation}

where $a=(d^{2}\langle N(E_{F})v_{x}^{2}\rangle /dE_{F}^{2})/12$, $b=\langle
N(E_{F})v^{-1}\rangle /2$, $v_{F}=\sqrt{3(d^{2}\langle
N(E_{F})v_{x}^{2}\rangle }$,  $Q=q_{c}\sqrt{a/bv_{F}},$ and $q_{c}$ is the
cutoff parameter for integration in Eq.\ref{FDT}. The physical meaning of
these parameters is as follows. $a$ defines the rate at which the static
susceptibility $\chi (q,0)$ falls away from the zone center, {\em i.e.}
the extent to which the tendency to ferromagnetism is stronger than that to
antiferromagnetism. This translates into the phase space in the Brillouin
zone where the spin fluctuations are important. $b$ controls the dynamic
effects in spin susceptibility.
The cutoff parameter $q_{c}$ is the least well defined quantity in this
formalism. One obvious choice is $q_{c}=$ $\sqrt{N(E_{F})/a},$ because for
larger $q$ the approximation (\ref{ab}) gives unphysical negative values for
the static susceptibility. On the other hand, this choice leads to 
noticeably different cutoffs for the two compounds, while 
one may argue that $q_c$ should
reflect mainly the geometry of the Fermi surface 
and thus be practically the same in these two cases. 
Furthermore, the fermiology of these compounds is very complicated:
in the paramagnetic state, there are 4 Fermi surfaces, two small and two
large (one open and one closed). In this situation, it is hardly
possible to justify any simple prescription for $q_c$. Therefore, we
have chosen a different route: we assume that $q_c$ is the {\it same} for
both materials, and choose a number which yields a good description of
both the equilibrium moment in Ni$_3$Al and the paramagnetic susceptibility
in Ni$_3$Ga, $q_c=0.382$ $a_0^{-1}$. Note that this is larger that the diameters
of the small Fermi surfaces but smaller than the radius of the Brillouin
zone, $\approx 0.5$ $a_0^{-1}$.

To calculate the above quantities, especially $a,$ we need accurate values
of the velocities on a fine mesh. Numerical differentiation of energies
within the tetrahedron method proved to be too noisy. Therefore we use the
velocities obtained analytically as matrix elements of the momentum
operator, computed within the \textit{optic} program of the WIEN package. A
bootstrap method\cite{boot},
as described in Ref.\cite{larson}, was used to obtain
stable values for $a,b$. We found for Ni$_{3}$Al,
using as the energy unit Ry, the length unit Bohr, and
the velocity unit Ry$\cdot $Bohr,
$a=230$, $b=210$, $v_{F}=0.20$,  and $\xi =0.445$ $\mu_B$.
For Ni$_{3}$Ga $a=140$, $b=270$, $v_{F}=0.19$,  and $%
\xi =0.556$ $\mu_B$. Using the resulting values of $\xi $
each compound we obtain a magnetic moment of $M=0.3$ $\mu_B $/cell for Ni$%
_{3}$Al and a paramagnetic state with the renormalized susceptibility $\chi
(0,0)=1/\tilde{a}_{2}=6.8\times10^{-5}$ emu/g for Ni$_{3}$Ga, thus correcting
the incorrect ordering of the magnetic tendencies of these two compounds
and reproducing extremely well the experimental numbers of $M=0.23$ $\mu_B$, $\chi
(0,0)=6.7\times 10^{-5}$ emu/g, respectively.
This qualitative behavior is due to
the different coefficient $a,$ {\em i.e.}, different $q$
dependencies of $\chi _{0}(q,0)$ at small $q$,
which relates to the phase space available for soft fluctuations.

Now we turn to the pressure dependence. The above results imply that
beyond LDA fluctuations are already larger than the moments themselves
at $P=0$. In this regime, we may assume that the size of the
beyond LDA fluctuations is only weakly pressure dependent.
Then we can apply Eq.
\ref{corr} to the data shown in Fig. \ref{P}
using $\xi=0.47$ as needed to match the $P=0$ value of $M$.
This yields a value $P_c$=10 GPa in quite good agreement with
the experimental value, $P_c$=8.1 GPa.
\cite{niklowitz}

In conclusion, we address the LDA failure to describe the physics of 
magnetism in Ni$_3$Al and Ni$_3$Ga even qualitatively. We identify
the problem as neglect of spin fluctuations
associated with the ferromagnetic quantum critical point.
These are stronger in Ni$_3$Ga despite the fact that the latter
has a larger density of states and is therefore  
more magnetic in mean-field theories. The reason for the difference
in the spin fluctuation spectra is in
the {\bf q} dependence of the non-interacting 
spin susceptibility.

We are grateful for helpful discussions with
G.G. Lonzarich and S.S. Saxena.
Work at the Naval Research Laboratory is supported by the Office of
Naval Research.

\begin{table}[tbp]
\caption{Magnetic energy (see text), magnetic moment in $\protect\mu _{B}$/cell
and $N(E_{F})$ in eV$^{-1}$ on a per spin per formula unit basis.}
\begin{tabular}{ccccc}
  & $|\Delta E|$ (meV)&$M$ (calc.)&$M$(expt.)&  $N(E_F)$ \\
\tableline
Ni$_3$Al & 10.3 & 0.71  & 0.23  &  3.2  \\
Ni$_3$Ga & 14.3 & 0.79  & 0.00  &  3.4  \\
\end{tabular} 
\label{tab:table1} \end{table}
\end{multicols}


\begin{references}

\bibitem{laughlin}
R.B. Laughlin,
G.G. Lonzarich, P. Monthoux and D. Pines,
Adv. Phys. {\bf 50}, 361 (2001).

\bibitem{saxena} S.S. Saxena,
P. Agarwal, K. Ahilan, F.M. Grosche, R.K.W.
Haselwimmer, M.J. Steiner, E. Pugh, I.R. Walker, S.R. Julian, P. Monthoux,
G.G. Lonzarich, A. Huxley, I. Sheikin, D. Braithwaite, and J. Flouquet,
Nature {\bf 406}, 587 (2000).

\bibitem{aoki} D. Aoki,
 A. Huxley, E. Ressouche, D. Braithwaite, J. Flouquet,
J.P. Brison, E. Lhotel, and C. Paulsen, 
Nature {\bf 413}, 613 (2001).

\bibitem{pfleiderer} C. Pfleiderer,
M. Uhlarz, S.M. Hayden, R. Vollmer,
H. von Lohneysen, N.R. Bernhoeft, and G.G. Lonzarich, 
Nature {\bf 412}, 58
(2001).

\bibitem{pf2}
C.P. Pfleiderer, S.R. Julian and G.G. Lonzarich, Nature {\bf 414}, 427 (2001).

\bibitem{boer} F.R. de Boer,
C.J. Schinkel, J. Biesterbos, and S. Proost, 
J. Appl. Phys. \textbf{40}, 1049 (1969).

\bibitem{niklowitz}
P.G. Niklowitz, F. Beckers, N. Bernhoeft, D. Braithwaite,
G. Knebel, B. Salce, J. Thomasson, J. Floquet and G.G. Lonzarich
(unpublished); presented at Conference on Quantum Complexities in
Condensed Matter, 2003.
\bibitem{hayden} S.M. Hayden, G.G. Lonzarich and H.L. Skriver,
Phys. Rev. B \textbf{33}, 4977
(1986).

\bibitem{steiner}
M.J. Steiner, F. Beckers, P.G. Nicklowitz and G.G. Lonzarich,
Physica B {\bf 329}, 1079 (2003).

\bibitem{sr-327} D.J. Singh and I.I. Mazin, Phys. Rev. B
\textbf{63}, 165101
(2001)

\bibitem{zrzn2} D.J. Singh and I.I. Mazin, Phys. Rev. Lett. \textbf{88},
187004 (2002).

\bibitem{sc3in} A. Aguayo and D.J. Singh, Phys. Rev. B \textbf{66}, 020401
(2002).

\bibitem{nac2}
D.J. Singh, Phys. Rev. B {\bf 68}, 020503 (2003).

\bibitem{hubbard}
This is distinct from
the well-known problems that the LDA has in systems with strong Hubbard
correlations, so such a signature should not be interpreted as evidence
of strong Hubbard correlations.

\bibitem{buiting} J.J. Buiting, J. Kl\"ubler, and F.M. Mueller, 
J. Phys. F {\bf 39} L179 (1983).

\bibitem{moruzzi1} V.L. Moruzzi and P.M. Marcus, Phys. Rev. B,
{\bf 42}, 5539  (1990).

\bibitem{min} B.I. Min, A.J. Freeman, and H.J.F. Jansen, Phys. Rev. B 
\textbf{37}, 6757 (1988).

\bibitem{xu} J.H. Xu,
B.I. Min, A.J. Freeman, and T. Oguchi, 
Phys. Rev. B 
\textbf{41}, 5010 (1990).

\bibitem{guo} G.Y Guo, Y.K. Wang, Li-Shing Hsu, J. Magn. Magn. Mater.
\textbf{239}, 91 (2002).

\bibitem{hsu}
L.-S. Hsu, Y.-K. Wang and G.Y. Guo, J. Appl. Phys. {\bf 92}, 1419 (2002).

\bibitem{singh-book}
D.J. Singh, {\em Planewaves Pseudopotentials and the LAPW Method}
(Kluwer Academic, Boston, 1994).

\bibitem{singh-lo} D. Singh, Phys. Rev. B 
{\bf 43}, 6388 (1991).

\bibitem{wei} S.H. Wei and H. Krakauer, Phys. Rev. Lett.  {\bf 55}, 1200 (1985).

\bibitem{WIEN} P. Blaha, K. Schwarz G.K.H. Madsen, D. Kvasnicka, and J.
Luitz, \textit{WIEN2K, An Augmented Plane Wave + Local Orbitals Program for
for Calculating Crystal Properties} (K. Schwarz, Techn. Universitat Wien,
Austria, 2001), ISBN 3-9501031-1-2.

\bibitem{hl} L. Hedin and B. Lundqvist, J. Phys. C, {\bf 4}, 2064 (1971).
\bibitem{hl2}U. von Barth and L. Hedin, J. Phys. C {\bf 5}, 1629 (1972).

\bibitem{Krasko} G.L. Krasko, Phys. Rev. B, {\bf 36} 8565 (1987).

\bibitem{Pnote} To compute pressure, we used
$P=B/B'[(V/V_0)^{B'}-1]$, where $V/V_0$ is the volume compression,
$B$ and $B'$ are
the bulk modulus and its derivative. We used the experimental bulk
modulus of Ni$_3$Al, $B$=174 GPa, \cite{wallow} which agrees with the
experimental volume
LDA value. \cite{osburn}
For $B'$ we used the calculated value $B'$=5.2.

\bibitem{shimizu} M. Shimizu, Rep. Prog. Phys. \textbf{44}, 329 (1981).

\bibitem{moriya} T. Moriya and A. Kawabata, J. Phys. Soc. Jpn.
{\bf 34}, 639 (1973).

\bibitem{larson} P. Larson, I.I. Mazin and D.J. Singh, cond-mat/0305407.

\bibitem{boot}B. Efron and R.J. Tibshirani, {\em
An Introduction to the Bootstrap} (Chapmann and Hall, New York, 1993).

\bibitem{schinkel} C.J. Schinkel, F.R. de Boer, and B. de Hon, J.
Phys. F \textbf{3}, 1463 (1973).

\bibitem{wallow}
F. Wallow, G. Neite, W. Schroer and E. Nembach, 
Phys. Status Solidi A {\bf 99}, 483 (1987).

\bibitem{osburn}
J.E. Osburn, M.J. Mehl and B.M. Klein, Phys. Rev. B {\bf 43}, 1805 (1991).

\end{references}
\end{document}